\documentclass[showpacs,preprint,preprintnumbers,amsmath,showkeys, amssymb, graphicx, graphics, epsfig]{revtex4}
\input{epsf}

\usepackage{epsfig}
\usepackage{dcolumn}
\usepackage{bm}

\begin{document}

\def\be{\begin{equation}}
\def\ee{\end{equation}}
\def\bdm{\begin{displaymath}}
\def\edm{\end{displaymath}}
\def\erfc{\hbox{erfc }}
\def\vpa{v_{\parallel }}
\def\vper{v_{\perp }}
\def\Omm{\Omega }
\def\ppa{p_{\parallel }}
\def\pper{p_{\perp }}
\def\ppv{\vec{p}}
\def\kkv{\vec{k}}
\def\omm{\omega}
\def\krz{{\bf \times }}
\def\dele{\delta \vec{E}(\vec{k},\omega )}
\def\delb{\delta \vec{B}(\vec{k},\omega )}
\def\Ab{\sum_a\omm _{p,a}^2(m_ac)^3}
\def\erf{\rm {erf }}
\title{Cumulative effect of Weibel-type instabilities in counterstreaming
plasmas with non-Maxwellian anisotropies}

\author{M. Lazar \footnote{Electronic mail: mlazar@tp4.rub.de}, R. Schlickeiser
and P. K. Shukla}
\affiliation{Institut f\"ur Theoretische Physik, Lehrstuhl IV:
Weltraum- und Astrophysik, Ruhr-Universit\"at Bochum, D-44780 Bochum, Germany }

\date{\today}

\begin{abstract}

Counterstreaming plasma structures are widely present in laboratory experiments 
and astrophysical systems, and they are investigated either to prevent 
unstable modes arising in beam-plasma experiments or to prove the existence of 
large scale magnetic fields in astrophysical objects.
Filamentation instability arises in a counterstreaming plasma
and is responsible for the magnetization of the plasma. 
Filamentationally unstable mode is described by assuming that 
each of the counterstreaming plasmas has an isotropic 
Lorentzian (kappa) distribution. In this case, the filamentation instability
growth rate can reach a maximum value markedly larger than that for a 
a plasma with a Maxwellian distribution function. This behaviour is opposite 
to what was observed for the Weibel instability growth rate in a bi-kappa 
plasma, which is always smaller than that obtained for a bi-Maxwellian plasma.
The approach is further generalized for a counterstreaming plasma with a 
bi-kappa temperature anisotropy. In this case, the filamentation instability
growth rate is enhanced by the Weibel effect when the plasma is hotter 
in the streaming direction, and the growth rate becomes even larger. 
These effects improve significantly the efficiency of the magnetic field 
generation, and provide further support for the potential role of 
the Weibel-type instabilities in the fast magnetization scenarios.

\end{abstract}


\maketitle

The electromagnetic instabilities of Weibel-type are driven by the 
velocity anisotropy of the plasma particles, whether it resides in a 
plasma with an electron temperature anisotropy (Weibel instability \cite{w59, zm07}),
or in a counterstreaming plasma system (filamentation instability \cite{f59}).
Davidson \cite{d83}, has combined for the first time these two sources of free
energy showing that the Weibel-type instabilities release the excess of 
free energy stored in "the relative directed motion between plasma 
components and/or an anisotropy in plasma kinetic energy" \cite{d83}.
Moreover, in order to describe a counterstreaming plasma with temperature anisotropies, 
Davidson \cite{d83} defined a new counterstreaming distribution function  
including a bi-Maxwellian electron temperature distribution function, and  
derived a dispersion relation, which admits purely growing solutions 
propagating perpendicular to the electron streaming direction. 
In a counterstreaming plasma with electron temperature anisotropies, 
the both free energy sources are present. 
Their contributions cumulate leading either to 
enhancing or quenching the electromagnetic instability,
depending on the plasma kinetic energy perpendicular to the 
propagation direction of the instability, if it exceeds or not 
the parallel kinetic energy \cite{bd06, lss06, sl08}. 

The stability properties of the counterstreaming plasma have been 
extensively investigated \cite{bd06, lss06, sl08, ct81, y93, kj95, s02, sd03, ts05a},
and the general observation is that the effect of a finite plasma temperature
has been revealed only for a Maxwellian electron distribution function.
Further investigations have been carried out in order to examine these instabilities 
in a magnetized counterstreaming plasma with a bi-Maxwellian electron temperature 
anisotropy \cite{l71, a73, ts05b}. As expected, the presence of a dc magnetic field 
in the streaming direction inhibits the Weibel-type instabilities \cite{ct81, hd05, ts07}.
We therefore ignore any influence of an ambient magnetic field in the present investigation.

Space plasmas are observed to possess a non-Maxwellian particle 
distribution function with a high energy tail \cite{v68,lk83}. 
The most common distribution function used to model 
such natural plasmas is the generalized kappa distribution \cite{st91}

\be
f_{\kappa} (v) = {1 \over \pi^{3/2} \theta^3} \,
{\Gamma (\kappa + 1) \over \kappa^{3/2} \Gamma(\kappa -1/2)} \, 
\left(1+{v^2 \over \kappa \theta^2} \right)^{-\kappa -1}, \label{e1} 
\ee
where $\kappa$ is the spectral index (for typical space plasmas, $\kappa$ 
generally lies in the range 2--6), and thermal velocity $\theta$ is related
to the particle temperature by $\theta^2 = \left[(2 \kappa -3) / \kappa \right] v_T^2$ and
$ v_T^2= (k_B T) / m$. The Maxwellian and kappa distributions differ substantially in the 
high energy tail ($|v/\theta| \gg 1$), but in the limit $\kappa \to + \infty $, $f_{\kappa}$ degenerates into a Maxwellian distribution function. The above $\kappa$-distribution has been used to explain different 
waves and instabilities and their related effects in isotropic plasmas where 
the velocity distribution function is assumed to deviate from the Maxwellian thermal equilibrium
due to the presence of high energy tails \cite{mpl97, hmv00}.

Very often plasma particles have a preferred direction of motion in space 
(e.g. a high temperature direction, a stationary magnetic field direction, 
or a streaming direction) and their velocity distribution function is anisotropic. 
In order to describe the anisotropic plasmas with high-energy tails, one
can use a two-temperature bi-kappa distribution function \cite{st91}

\be
f_{\kappa} (v_x, v_y, v_z) = {1 \over  \pi^{3/2} \theta^2 \theta_{y}} \,
{\Gamma (\kappa + 1) \over \kappa^{3/2} \Gamma(\kappa -1/2)} \, 
\left(1+{v_x^2 + v_z^2 \over \kappa \theta^2} +
{v_{y}^2 \over \kappa \theta_{y}^2} \right)^{-\kappa -1}, \label{e2} 
\ee
which approaches the bi-Maxwellian distribution function as $\kappa \to + \infty $.
The temperature anisotropy is defined as $A= ({T_y / T})-1$, with

\be
\theta^2 = \left(2 \kappa -3 \over \kappa \right) v_{T_x}^2; \;\;\; \theta_y ^2=\left(2 \kappa -3 \over \kappa \right) v_{T_y}^2, \label{e3}
\ee
and
\be
v_{T_x}^2=v_{T_z}^2 = {k_B T \over m}; \;\;\;  v_{T_y}^2 = {k_B T_y \over m}. \label{e4}
\ee

The bi-kappa distribution function has been earlier attributed to 
magnetized plasmas providing a useful tool for a more realistic characterization
of the turbulent fields and acceleration mechanisms in the planetary environment \cite{l83, st92, m98}.
Moreover, Hellberg and Mace (2002) have generalized the plasma dispersion function for 
a kappa-Maxwellian distribution, and investigated the electrostatic waves arising
in such an anisotropic plasma. The kappa-Maxwellian distribution function combines a
one-dimensional $\kappa$ distribution along a preferred direction in space, and a 
two-dimensional Maxwellian in the plane perpendicular to this direction. 
This choice is justified physically by the fact
that in many instances there is equilibration and isotropization in the perpendicular plane,
but there is preferential acceleration, leading to a power-law distribution along another preferred
direction \cite{hm02}.

Recently, a bi-kappa distribution function of form (\ref{e2}) has been 
used to evaluate the Weibel instability rates in an unmagnetized plasma \cite{zm07}.
For small values of $\kappa$, the growth rate is reduced, but as $\kappa$ 
is increased, the growth rate enhances approaching the results obtained with
a bi-Maxwellian distribution. However, it has already been shown \cite{bd06, lss06, sl08} 
that, in a counterstreaming plasma with bi-Maxwellian temperature anisotropies, the 
(electro)magnetic instabilities of the Weibel-type can be markedly faster when 
the plasma is hotter along the streaming direction, or they can be stabilized by
a higher transverse plasma temperature.
Moreover, there are many space plasma systems in which colliding plasma shells 
can be described as counterstreams (e.g. relative motion of fully ionized 
gaseous matters in intergalactic medium, interaction of internal or external 
fireballs with surrounding interstellar medium in gamma-ray burst sources, 
flares in the solar winds). We are therefore entitled to extend here the investigation
on the aperiodic electromagnetic instabilities in counterstreaming
plasmas with non-Maxwellian anisotropies.

\begin{figure}[h] \centering
    \includegraphics[width=60mm, height=45mm]{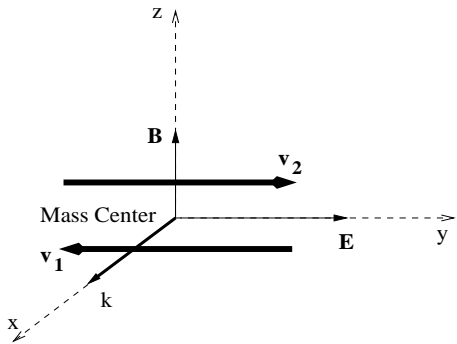}
\caption{Two counterstreaming plasmas and the electromagnetic 
filamentation mode with fields ${\bf E}$ and ${\bf B}$, 
and the wavevector ${\bf k}$ perpendicular to the direction of the streams.} \label{fig1} 
\end{figure}

In Figure \ref{fig1} we fix the orientation for the
counterstreams and the filamentation mode we are looking for: 
the electric field, ${\bf E} \parallel {\bf v_{1,2}}$, 
is along the streaming direction, and the wave vector, 
${\bf k} \perp {\bf v_{1,2}}$, is perpendicular to the streams.
This electromagnetic mode will be solution of the dispersion 
relation
\be
{k^2c^2 \over \omega^2} = \epsilon_{yy}, \label{e5}
\ee
where $\omega$ and $k$ are respectively, the frequency and
the wave number of the plasma modes, $c$ is the speed of light in vacuum, 
and $\epsilon_{yy}$ is the 
dielectric tensor which describes the oscillatory properties of the plasma 
system. From the linearized Vlasov-Maxwell equations, one simply obtains \cite{kmq68}

\be
\epsilon_{yy} =1-  \sum_a {\omm_{p,a}^2 \over \omega^2 }
\left[1 - k \int_{-\infty}^{+ \infty} \, d{\bf v} \,
{v_y^2 \over \omm - k v_x} \; {\partial f_{a,0} \over \partial v_x} \right], \label{e6}
\ee
where the unperturbed velocity distribution function $f_{a0}({\bf v})$ 
for the particles of species $a$ is considered normalized by $\int \, d{\bf v} \, f_{a0}({\bf v}) = 1$, 
and $\omm_{p,a} = (4 \pi n_a e^2 /m_a)^{1/2}$ is the plasma frequency.

We now introduce the distribution function 

\bdm
f_{\kappa} = {1 \over 2 \pi^{3/2} \theta^2 \theta_{y}} \,
{\Gamma (\kappa + 1) \over \kappa^{3/2} \Gamma(\kappa -1/2)} \, 
\left\{\left[1+{v_x^2 + v_z^2 \over \kappa \theta^2} +
{(v_{y}-v_0)^2 \over \kappa \theta_{y}^2} \right]^{-\kappa -1} \right. \edm 
\be \left. + \left[1+{v_x^2 + v_z^2 \over \kappa \theta^2} +
{(v_{y}+v_0)^2 \over \kappa \theta_{y}^2} \right]^{-\kappa -1} \right\}, \label{e7} 
\ee
which describes the electron plasma counterstreams with bi-Kappa temperature anisotropies.
This distribution function has to be added to the list of Lorentzian-type particle distribution
functions provided in Table I of Ref. \cite{st91}.
For the sake of simplicity, we assume symmetric counterstreams 
with equal densities, $\omega_{p,e,1}= \omega_{p,e,2} = \omega_{p,e}$, 
and equal temperature anisotropies, $A_1 = A_2= A= ({T_y / T})-1$ , 
and moving with the same velocities, $|v_1| = v_2 = v_0$.
Moreover, we assume that imobile ions form the neutralizing background, and the electron 
plasma counterstreams are considered homogeneous, and charge and current neutralized.

Inserting (\ref{e7}) into (\ref{e6}) and the resultant expression into (\ref{e5}) we derive the dispersion relation
\be
{k^2c^2 \over \omega^2} =\epsilon_{yy}^0 +  2 {\omm_{p,e}^2 \over \omega^2 }\; {v_0^2 \over \theta^2} \left[(1 - {1\over 2 \kappa})
+  {\omega \over k \theta} \, Z_{\kappa} \left({\omega \over k \theta} \right) \right]\, , \label{e8}
\ee
where 

\be
\epsilon_{yy}^0 = 1 + {\omm_{p,e}^2 \over \omega^2 }\; \left[A + (A+1)\, {\omega \over k \theta} \, Z_{\kappa}^0 \left({\omega \over k \theta}  
\right) \right] \label{e9}
\ee
has been obtained in Ref. \cite{zm07} for an anisotropic plasma with a bi-kappa distribution 
function without $v_0$, where the simple Weibel instability arises.
Otherwise, for $A= 0$ Eqs. (\ref{e8})--(\ref{e9}) simplify to

\be
{k^2c^2 \over \omega^2} =1 + {\omm_{p,e}^2 \over \omega k \theta}\; \, 
Z_{\kappa}^0 \left({\omega \over k \theta} \right) 
+ 2 {\omm_{p,e}^2 \over \omega^2 }\; {v_0^2 \over \theta^2} 
\left[(1 - {1\over 2 \kappa}) + {\omega \over k \theta} \, 
Z_{\kappa} \left({\omega \over k \theta} \right) \right], \label{e10}
\ee
which describes the filamentation instability arising from counterstreaming
plasmas with the same isotropic kappa distribution function.
The dispersion equation (\ref{e8}) contains the modified plasma dispersion function, 
defined in Ref. \cite{st91},
\be
Z_{\kappa}(f) = {1 \over \pi^{1/2} \kappa^{1/2}} \, {\Gamma (\kappa) \over \Gamma 
\left( \kappa -{1 \over 2}\right)} \, \int_{-\infty}^{+\infty} dx \, 
{(1+x^2/\kappa)^{-(\kappa +1)} \over x - f}, \; \;\; \Im (f) > 0 , \label{e11}
\ee
and that differs only slightly from the new modified plasma dispersion function 
\be
Z_{\kappa}^0(f) = {1 \over \pi^{1/2} \kappa^{1/2}} \, {\Gamma (\kappa ) \over \Gamma 
\left(\kappa -{1 \over 2}\right)} \, \int_{-\infty}^{+\infty} dx \, 
{(1+x^2/\kappa)^{- \kappa } \over x - f}, \; \;\; \Im (f) > 0 \label{e12}
\ee
introduced recently in Ref. \cite{zm07}. It is simple to demonstrate that the kappa 
dispersion functions from above are related to each other by

\be
Z_{\kappa}^0(f) = \left(1+{f^2 \over \kappa}\right) \, Z_{\kappa}(f) +
{f \over \kappa} \, \left(1-{1 \over 2k} \right), \nonumber
\ee
and therefore, both of them approach the plasma dispersion function \cite{fc61} 
in the limit of $\kappa \to \infty$. We should also remark that, in this limit, Eq. (\ref{e8})
becomes identical to the dispersion relation of cumulative filamentation-Weibel mode in 
a counterstreaming plasma with a bi-Maxwellian anisotropic distribution function 
(see Eq. (8) in Ref. \cite{sl08}).
It is clear now that the general dispersion relation given by 
(\ref{e8}) and (\ref{e9}), allows us to extend the investigation of purely 
growing filamentation instability by including the Weibel-type enhancing 
effects due to the existence of a non-Maxwellian temperature 
anisotropy in the counterstreaming plasmas. 

An analytical expression for the aperiodic solutions, 
$\Im(\omega) = \omega_i$, $\Re(\omega) =0$, of the general dispersion relations 
(\ref{e8}) and (\ref{e9}), can be further obtained only by
using the asymptotic forms of the modified plasma dispersion functions 
provided in Refs. \cite{zm07} and \cite{st91} for very large or very 
small arguments. Here, we keep the accuracy of a general approach and 
evaluate exactly the numerical growth rates for several representative cases.

\begin{figure}[h] \centering
    \includegraphics[width=60mm, height=60mm]{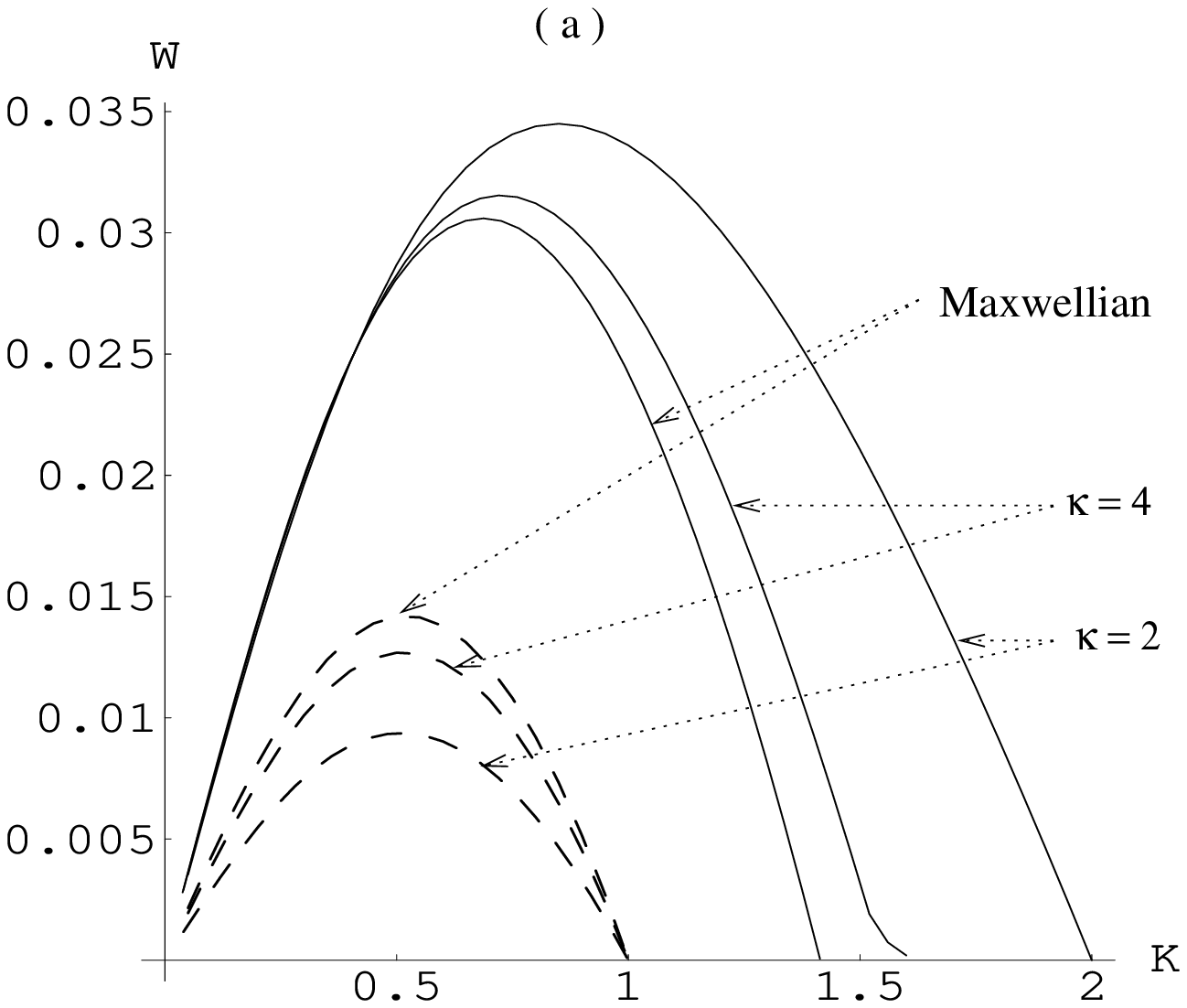}
    \includegraphics[width=60mm, height=60mm]{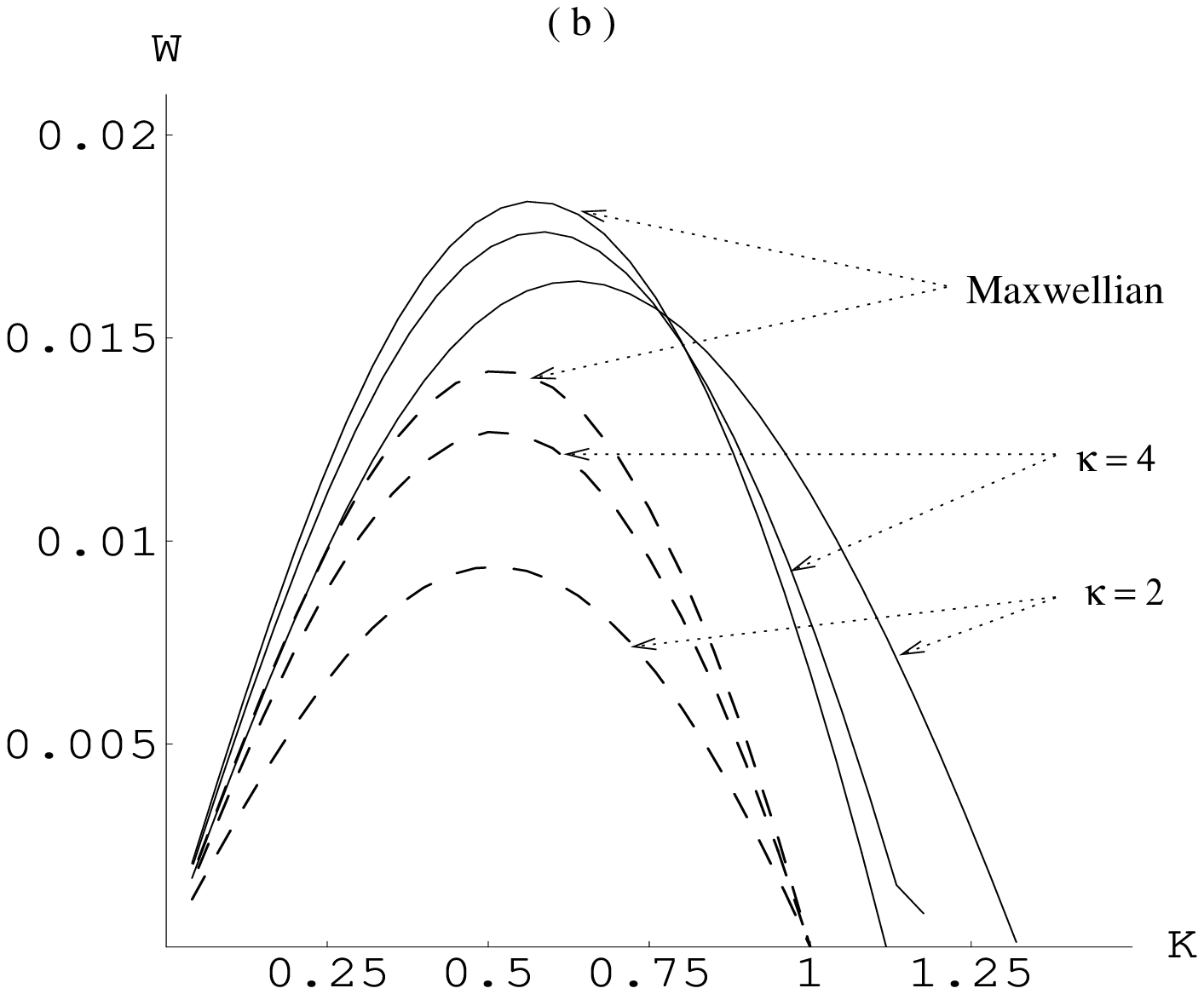}
\caption{The growth rates of the filamentation instability (solid lines) 
obtained from (\ref{e8}) for two values of relative streaming 
velocity: (a) $v_0/c= 0.1$ and (b) $v_0/c= 0.05$, and which 
cumulate the Weibel instability effect due to a bi-kappa 
temperature anisotropy, $A = 1$. The coordinates are scaled 
as W$ = \omega_i / \omega_{pe}$ and K$ = kc / \omega_{pe}$. 
For comparison, with dashed lines are shown the growth rates 
of simply Weibel modes ($v_0 = 0$).}\label{fig2} 
\end{figure}

\begin{figure}[h] \centering
    \includegraphics[width=60mm, height=60mm]{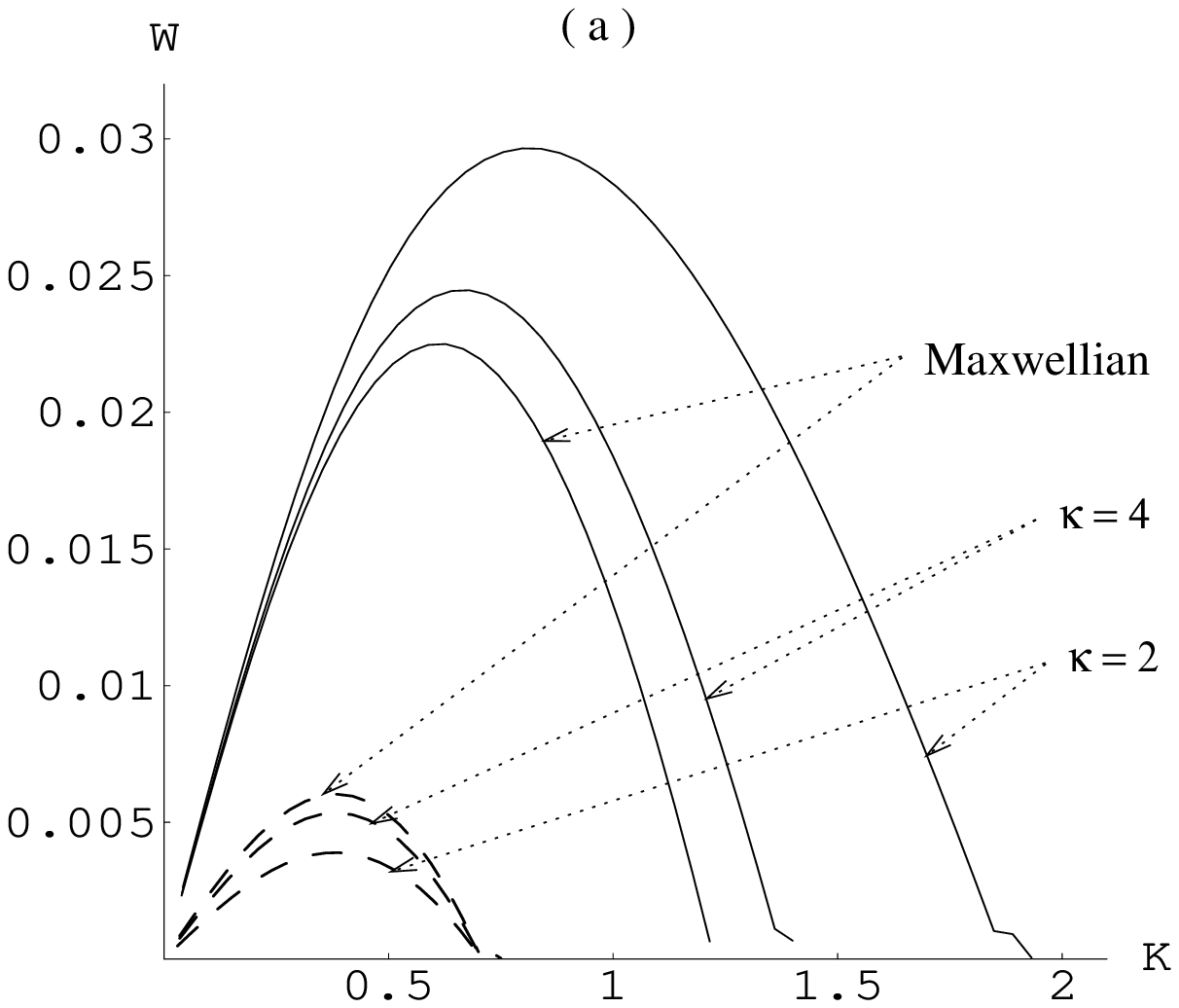}
    \includegraphics[width=60mm, height=60mm]{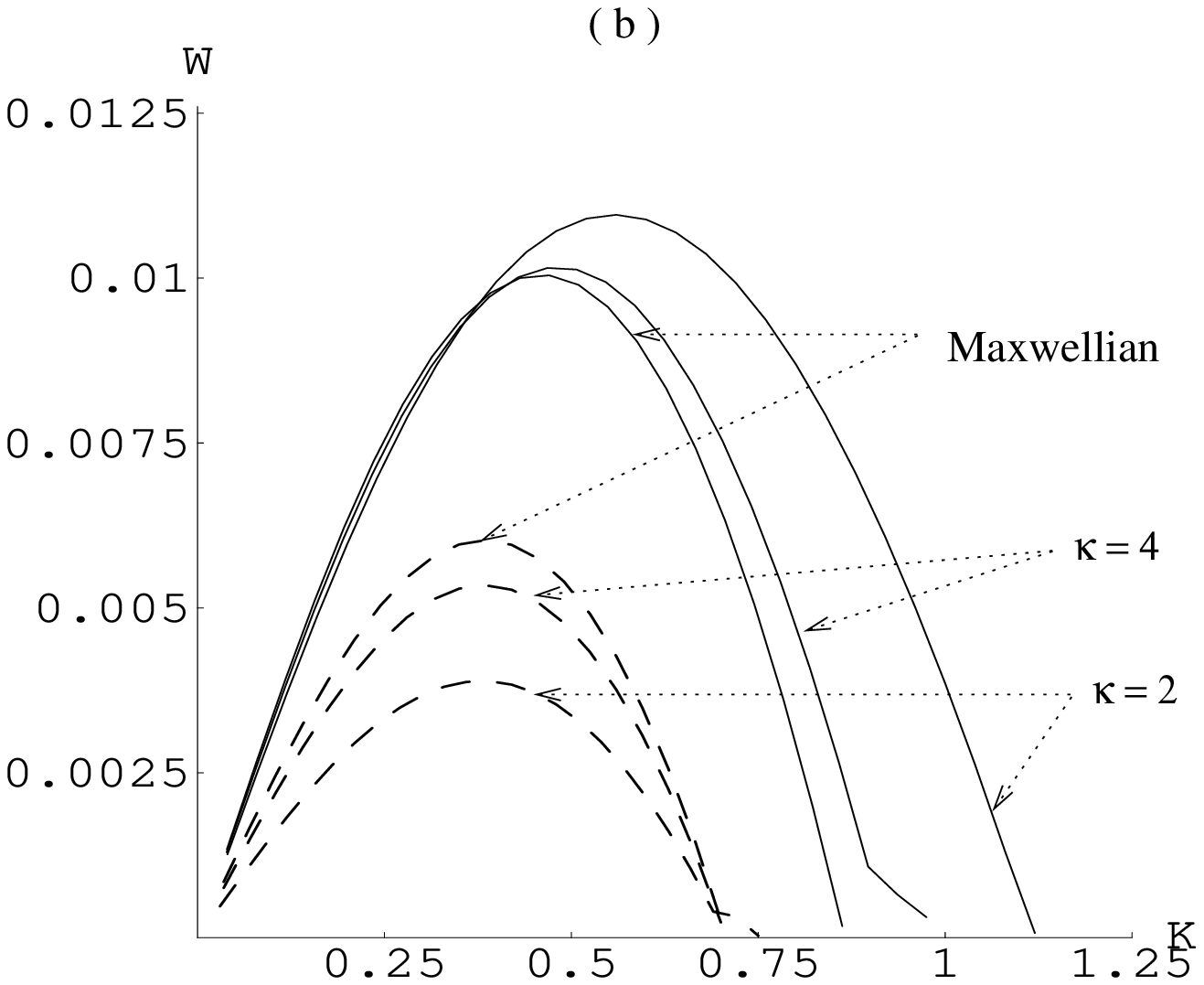}
\caption{The same as in Fig. \ref{fig2}, but for a smaller
anisotropy $A = 0.5$.} \label{fig3} 
\end{figure}

\begin{figure}[h] \centering
    \includegraphics[width=60mm, height=60mm]{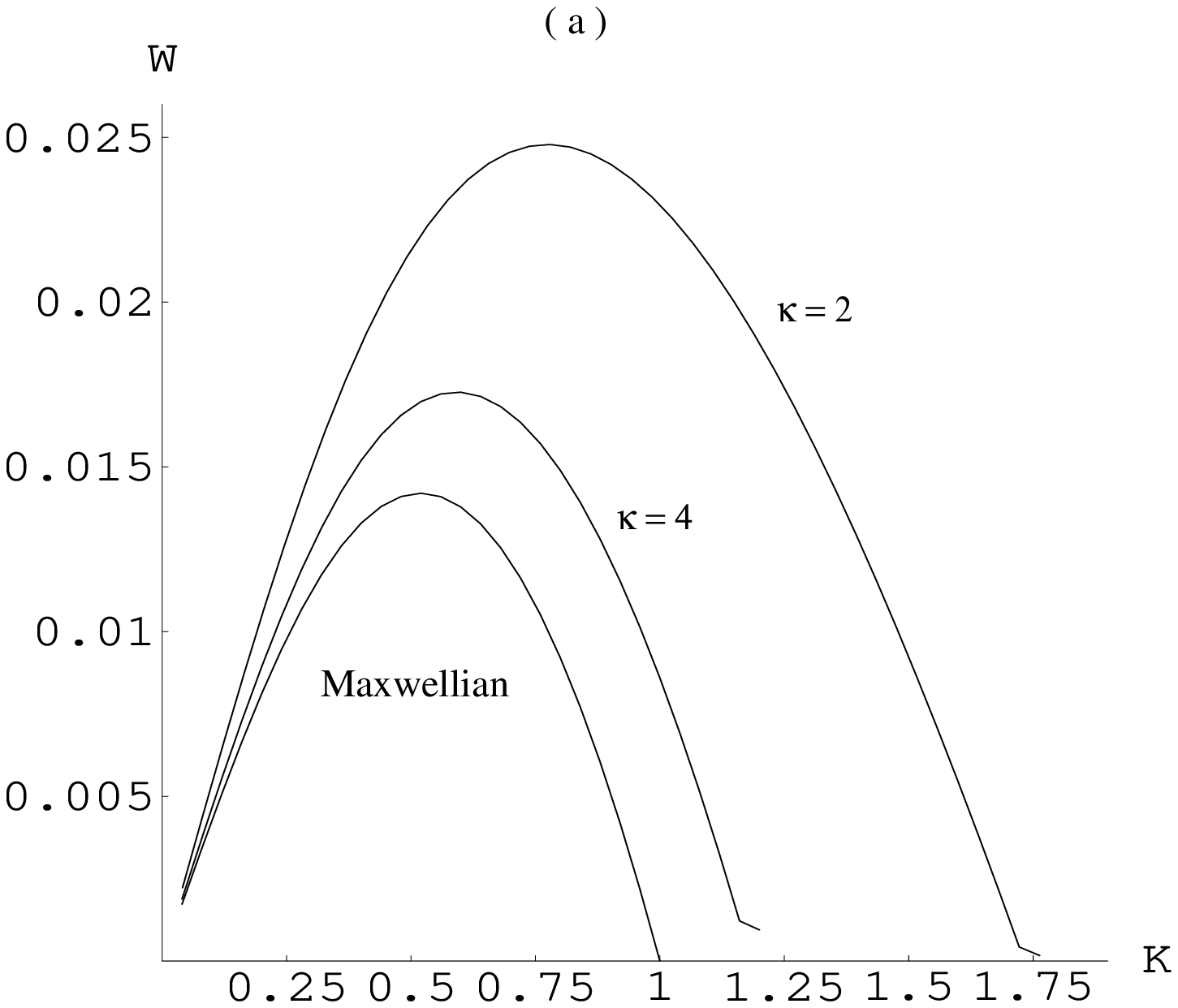}
    \includegraphics[width=60mm, height=60mm]{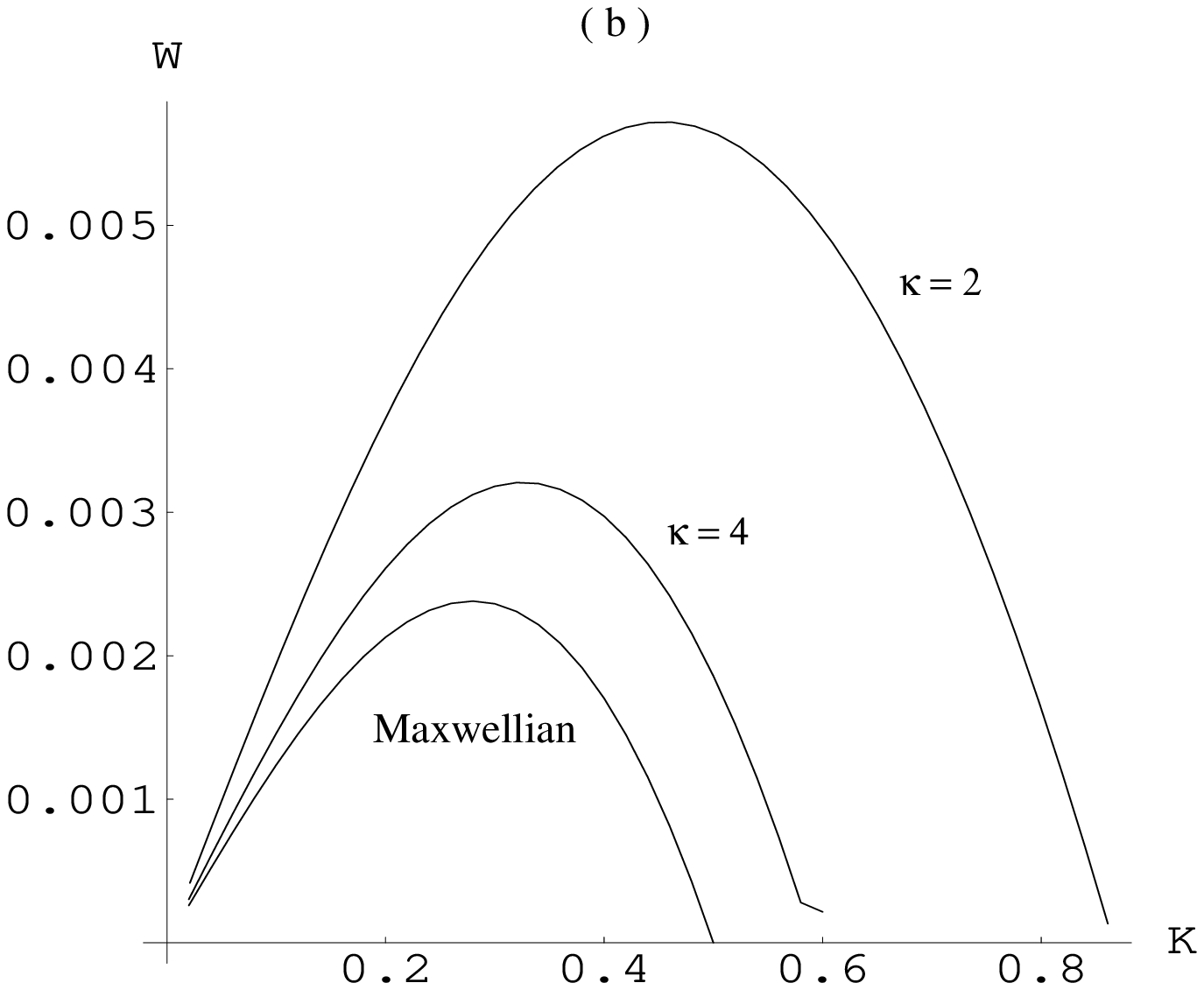}
\caption{The growth rates of the filamentation instability ($A=0$) 
obtained from (\ref{e10}) for two different values of relative streaming 
velocity: (a) $v_0/c= 0.1$ and (b) $v_0/c= 0.05$. On the ordinate and abscissa 
are W$ = \omega_i / \omega_{pe}$ and K$ = kc / \omega_{pe}$, respectively.}\label{fig4} 
\end{figure}

We first assume that the plasma is hotter in the streaming direction, e.g. $T_y > T$, that
means the anisotropy is positive. Thus, the aperiodic solutions
of Eq. (\ref{e8}), which correspond to the cumulative filamentation-Weibel 
mode, are plotted with solid lines in Fig. \ref{fig2} and Fig. \ref{fig3}
for two different values of the temperature anisotropies, $A = 1$ and 
$A = 0.5$, respectively. 
In Ref. \cite{zm07} it was shown that the growth rates of the
Weibel instability in a thermally anisotropic plasma with a bi-kappa
distribution function decrease in comparison to those obtained for the case
of a bi-Maxwellian electron distribution function. The Weibel growth rates
are plotted here with dashed line.
On the other hand, in counterstreaming plasmas with sufficiently large 
temperature anisotropies we observe the opposite effect: 
the growth rates of the cumulative filamentation-Weibel instability obtained
for bi-kappa anisotropies increase and become much higher than
those obtained for counterstreaming plasmas with bi-Maxwellian anisotropies. This is the 
case in Fig. \ref{fig2} (a) and Fig. \ref{fig3} (a),
provided a sufficiently large relative streaming velocity, $v_0$, is used.
Such behaviour is typical for the simple filamentations instability, 
which develops in isotropic plasma counterstreams ($A= 0$). 
In this case, the filamentation instability growth rates are solutions of 
Eq. (\ref{e10}), and they are shown in Fig. \ref{fig4} (a) and (b) for 
two different streaming velocities, $v_0/c= 0.1$ and $v_0/c= 0.05$, respectively.

Otherwise, if the streaming velocity, $v_0$, is small the kappa growth rates are 
lower than the Maxwellian, and this is shown in Fig. (\ref{fig2}) (b).

Due to a finite temperature in both plasma counterstreams 
the purely growing solutions appear to resemble a Weibel regime
and therefore their existence is limited to the small wave numbers less than
a cutoff value, $k \le k_c$, which is the nontrivial solution of (\ref{e8})
to the limit of $\omega_i = 0$

\be
k_c = {\omega \over c} \, \left[A + {v_0^2 \over \theta^2} \left( 2-{1\over \kappa}\right) \right]^{1/2}. \label{e13}
\ee

\begin{figure}[h] \centering
    \includegraphics[width=50mm, height=50mm]{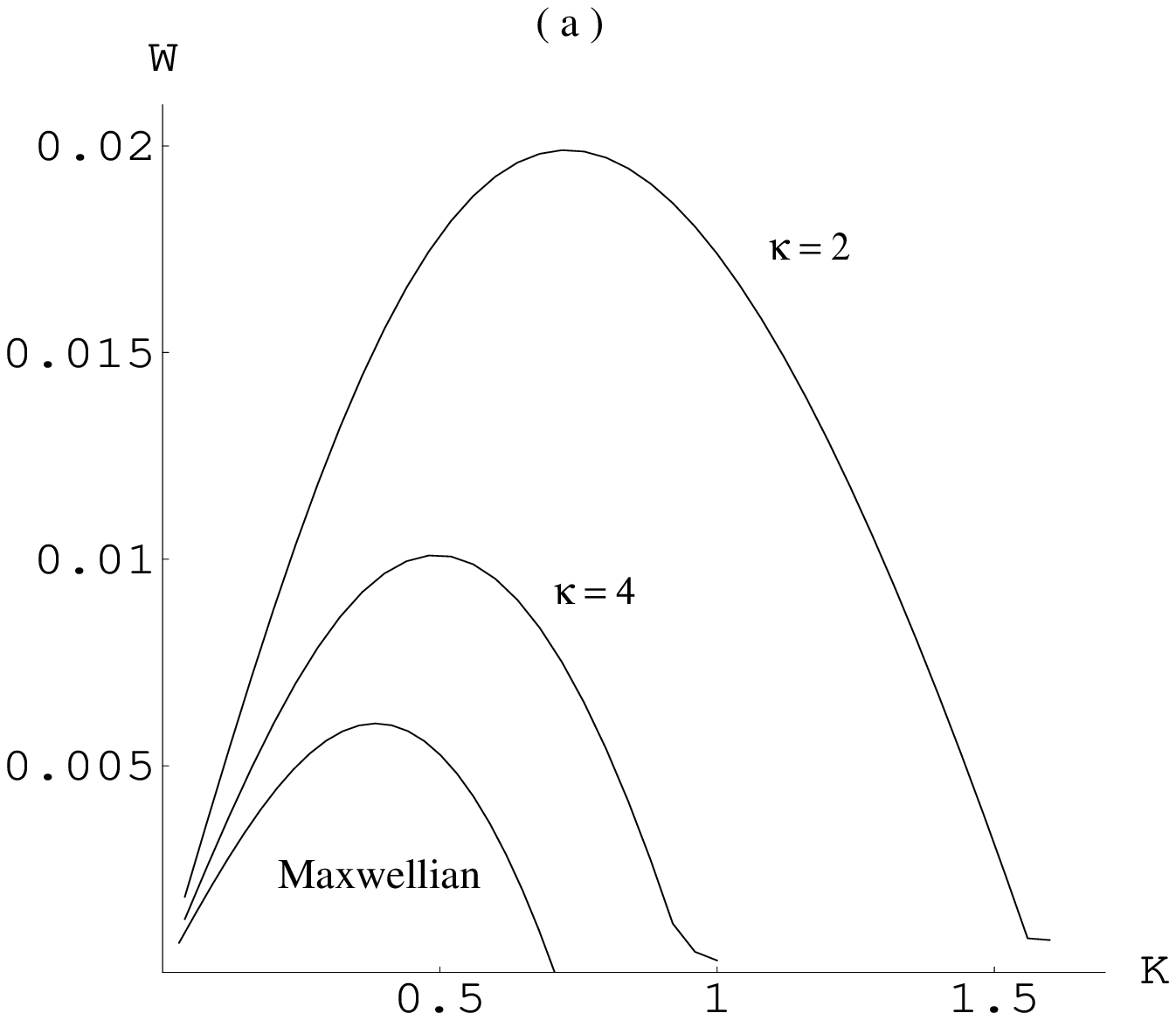}
    \includegraphics[width=50mm, height=50mm]{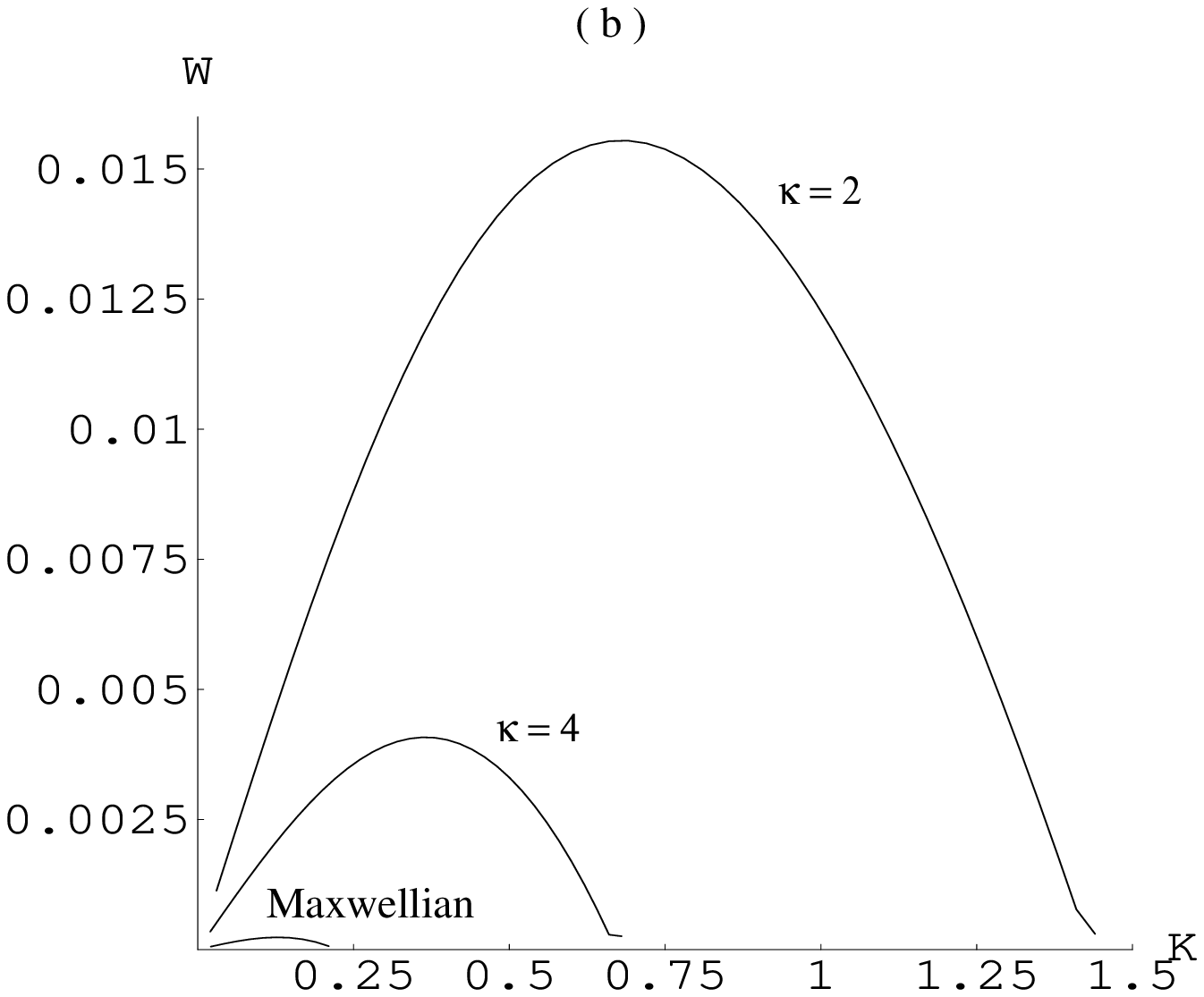}
    \includegraphics[width=50mm, height=50mm]{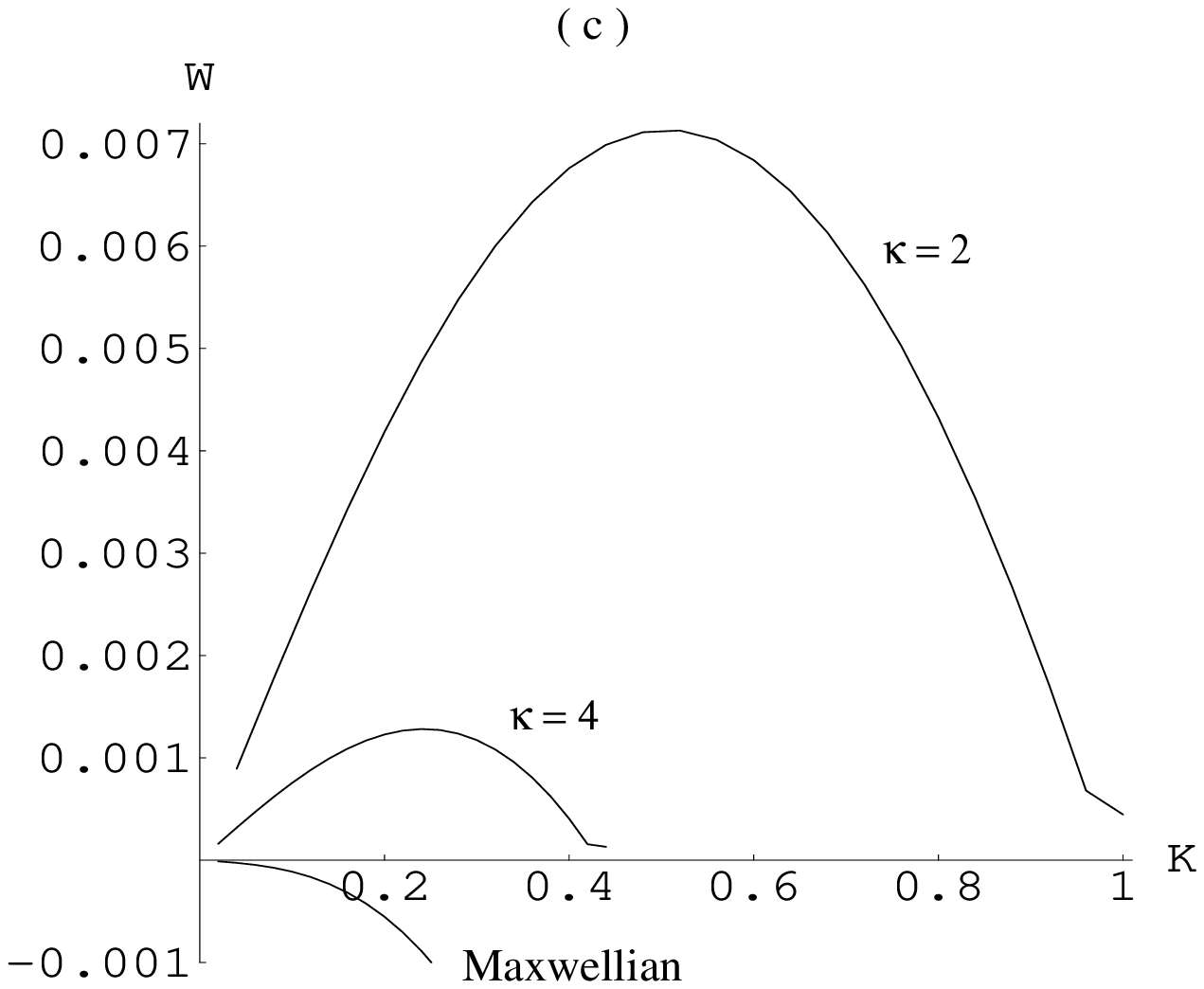}
\caption{The growth rates of the filamentation instability (solid lines) 
obtained from (\ref{e8}) for negative temperature anisotropies: 
(a) $v_0/c= 0.1$, $A= -0.5$, (b) $v_0/c= 0.1$, $A= -0.95$, and
(c) $v_0/c= 0.07$, $A= -0.5$. On the ordinate and abscissa 
are W$ = \omega_i / \omega_{pe}$ and K$ = kc / \omega_{pe}$, respectively.} \label{fig5} 
\end{figure}

Finally, when the contrastreaming plasma is colder in the streaming 
direction, i.e., $T_y < T$, the temperature anisotropy becomes negative, 
$ -1 < A < 0$. In this case, the effective velocity anisotropy imposed 
to the plasma particles by the relative motion of the counterstreaming plasmas, 
is reduced, and this explains the lower growth rates obtained 
in Fig. \ref{fig5} for the negative temperature anisotropies.
For a small streaming velocity, but a sufficiently large 
negative anisotropy, the instability is suppressed and 
instead natural damped modes occur. Further improvement 
of stabilization is observed for those modes corresponding 
to the larger kappa indices, which approach the Maxwellian modes.

To summarize, we have investigated the properties of purely growing electromagnetic 
instabilities in an anisotropic counterstreaming plasma and propagates perpendicular to 
the streaming direction. Specifically, we have considered for the first time 
the filamentation instability in counterstreaming plasmas with isotropic kappa distributions, 
and derived the dispersion relation (\ref{e10}), and displayed its numerical 
solutions in Fig. \ref{fig5}. We found that the aperiodic instability can reach maximum
growth rates significantly larger than those 
calculated for the plasma with a Maxwellian distribution function. This is
contrary to the behaviour of the Weibel instability in a 
bi-kappa plasma, which has growth rates smaller
than those obtained for a bi-Maxwellian plasma.

Moreover, we have generalized the theoretical approach assuming 
that each of the counterstreaming plasmas exhibits a temperature 
anisotropy of a non-Maxwellian type, namely, a
bi-kappa distribution function. In this case, the instability
is driven equally by the free energy stored in the relative motion
of counterstreaming plasmas and the temperature anisotropy of
the plasma particles. Comparing Figures \ref{fig2} and \ref{fig3}
with Figure \ref{fig5}, we observe that the filamentation instability
growth rate is enhanced in the presence of the positive temperature 
anisotropies, when plasma is hotter 
in the streaming direction. Higher growth rates are obtained for
larger temperature anisotropies. This is the first effect of the 
so-called cumulative filamentation-Weibel instability, and it improves
significantly the efficiency of the magnetic field generation, and
provides further support for the potential role of the Weibel-type
instabilities in the fast magnetization scenarios, e.g. gamma-ray
bursts, etc.

For negative temperature anisotropies the effective 
velocity anisotropy is reduced, and therefore 
the growth rates of the filamentation instability decrease, 
but they remain larger than those for the bi-Maxwellian plasmas. 
The unstable modes with largest kappa indices can be evenly
suppressed. This is the second cumulative effect of the 
counterstreaming plasmas and the (negative) temperature anisotropy, 
which contributes in this case to the stabilization of the plasma system.

\acknowledgements
{M.L. acknowledges financial support from the Alexander von Humboldt Foundation.
This work was partially supported by the 
Deutsche Forschungsgemeinschaft through the Sonderforschungsbereich 591.}


\begin{references}


\bibitem{w59}  E.S. Weibel, Phys. Rev. Lett. {\bf 2}, 83 (1959).
\bibitem{zm07} S. Zaheer and G. Murtaza, Phys. Plasmas {\bf 14}, 022108 (2007)
\bibitem{f59} B.D. Fried, Phys. Fluids {\bf 2}, 337 (1959).
\bibitem{d83} R.C. Davidson, in {\it Handbook of Plasma Physics}, edited by M.N. 
Rosenbluth and R.Z. Sagdeev (North-Holland, Amsterdam, 1983), Vol. 1, p. 519
\bibitem{bd06} A. Bret and C. Deutsch, Phys. Plasmas {\bf 13}, 022110 (2006)
\bibitem{lss06}	M. Lazar, R. Schlickeiser and P.K. Shukla, Phys. Plasmas {\bf 13}, 102107 (2006).
\bibitem{sl08}	A. Stockem and M. Lazar, Phys. Plasmas {\bf 15}, 014501 (2008).

\bibitem{ct81}  J.R. Cary, L.E. Thode, D.S. Lemons, M.E. Jones and M.A. Mostrom, Phys. Fluids {\bf 24}, 1818 (1981).
\bibitem{y93} T.-Y. B. Yang, Y. Gallant, J. Arons and A.B. Langdan,
Phys. Fluids {\bf B 5}, 3369 (1993).
\bibitem{kj95} J. Krall and G. Joyce, Phys. Plasmas {\bf 2}, 1326 (1995).
\bibitem{s02} L.O. Silva, R.A. Fonseca, J.W.  Tonge, W.B. Mori, and J.M. Dawson, Phys. Plasmas {\bf 9}, 2458 (2002).
\bibitem{sd03} E.A. Startsvev and R.C. Davidson, Phys. Plasmas {\bf 10}, 4829 (2003).
\bibitem{ts05a} R.C. Tautz and R. Schlickeiser, Phys. Plasmas {\bf 12}, 072101 (2005).
\bibitem{l71} K.F. Lee and J.C. Armstrong, Phys. Rev. A {\bf 4}, 2087 (1971).
\bibitem{a73} S.S. Aggarwal and S.P. Talwar, Astrophys. Space Sci. {\bf 23}, 315 (1973);
S.S. Aggarwal, Astrophys. Space Sci. {\bf 33}, 259 (1975)
\bibitem{ts05b} R.C. Tautz and R. Schlickeiser, Phys. Plasmas {\bf 12}, 122901 (2005); Phys. Plasmas {\bf 13}, 062901 (2006).
\bibitem{hd05} C.B. Hededal and K.-I. Nishikawa, Astrophys. J. {\bf 623}, L89 (2005).
\bibitem{ts07} R.C. Tautz and J.-I. Sakai, Phys. Plasmas {\bf 14}, 012104 (2007).

\bibitem{v68} V.M. Vasyliunas, J. Geophys. Res. {\bf 73}, 2839 (1968).
\bibitem{lk83} A.T.Y. Lui and S.M. Krimigis, Geophys. Res. Lett. {\bf 10}, 13 (1983).

\bibitem{st91} D. Summers and R.M. Thorne, Phys. Fluids {\bf B 3}, 1835 (1991).
\bibitem{mpl97} M. Maksimovic, V. Pierrard and J.F. Lemaire, Astron. Astrophys. {\bf 324}, 725 (1997).
\bibitem{hmv00} M.A. Hellberg, R.L. Mace and F. Verheest, in {\it Waves in Dusty, Solar and Space Plasmas, Leuven, 2000}, edited by F. Verheest, M. Goosens, M.A. Hellberg, and R. Bharuthram (AIP, Melville, NY, 2000), p. 348.

\bibitem{l83} M.P. Leubner, J. Geophys. Res. {\bf 88}, 469 (1983).
\bibitem{st92} D. Summers and R.M. Thorne, J. Geophys. Res. {\bf 97}, 16827 (1992).
\bibitem{m98} R.L. Mace, J. Geophys. Res. {\bf 103}, 14643 (1998).
\bibitem{hm02} M.A. Hellberg and R.L. Mace, Phys. Plasmas {\bf 9}, 1495 (2002). 

\bibitem{kmq68} G. Kalman, C. Montes and D. Quemada, Phys. Fluids 11, 1797 (1968).



\bibitem{fc61} B.D. Fried, \& S.D. Conte, The Plasma Dispersion Function, (Academic Press, New York 1961).


\end{references}
\end{document}